\documentclass[aps,prd,twocolumn,superscriptaddress,nofootinbib,floatfix]{revtex4-2}

\usepackage[T1]{fontenc}
\usepackage[utf8]{inputenc}
\usepackage{lmodern}
\usepackage[reqno]{amsmath}    
\usepackage{bm,amssymb,mathtools}
\usepackage{graphicx}
\usepackage{booktabs}
\usepackage{xcolor}
\usepackage[hypertexnames=false]{hyperref}
\hypersetup{colorlinks=true,linkcolor=blue,citecolor=blue,urlcolor=blue}

\usepackage{tikz}
\usetikzlibrary{arrows.meta,positioning,calc}
\usepackage{pgfplots}
\pgfplotsset{compat=1.18}

\newcommand{\e}{\epsilon}      
\newcommand{\Bpar}{B}          
\newcommand{\pow}{p}           
\newcommand{\ordone}{\mathcal{O}(1)}
\newcommand{\phiGR}{\varphi}   
\newcommand{\MS}{\overline{\text{MS}}}

\begin{document}

\title{Two-over-Two Lattice Flavor from a Single Flavon with Three Messenger Chains}

\author{Vernon Barger}
\affiliation{Department of Physics, University of Wisconsin--Madison, Madison, WI 53706, USA}

\date{Accepted February 19, 2026}

\begin{abstract}
Flavor hierarchies are organized by a single parameter $\Bpar\simeq 5.357$ in a single-flavon Froggatt--Nielsen (FN) framework, in which each effective Yukawa entry arises from the sum of \emph{three} unit-magnitude messenger chains.
We present benchmark complex $\ordone$ Yukawa matrices that reproduce quark and charged-lepton masses at $M_Z$ as powers of $\e\equiv 1/\Bpar$.
The organizing principle is a two-over-two (2/2) lattice of quadrilateral mass ratios, which maps directly to a rational lattice of FN exponents.
Sequential dominance preserves the leading-power exponent matrices, while subleading messenger chains generate entry-dependent complex $\ordone$ coefficients and provide a UV-friendly origin for CP violation.
Neutrino masses are discussed at the level of eigenvalues within the same $\Bpar^n$ counting.
\end{abstract}

\maketitle

\section{Introduction}
\label{sec:intro}

The fermion generation structure is one of the big mysteries of the Standard
Model (SM).  Quark and lepton masses span many orders of magnitude, and the
CKM and PMNS matrices contain small and large mixing angles as well as
CP-violating phases.  Despite the huge amount of experimental data, we still
lack a simple organizing principle that explains why the flavor parameters take
the values we observe.

One of the most economical ideas is the Froggatt--Nielsen (FN) mechanism~\cite{Froggatt:1978nt,Leurer:1992qj,Leurer:1993gy}.
In its simplest form, FN postulates a horizontal flavor symmetry (often a
$U(1)_{\text{FN}}$) that is spontaneously broken by the vacuum expectation
value (VEV) of a complex scalar ``flavon'' field $\Phi$.  Heavy messenger
fields at a high scale $\Lambda$ communicate this breaking to the SM fermions,
so that the effective Yukawa couplings are suppressed by powers of a single
small parameter
\begin{equation}
  \epsilon \equiv \frac{\langle \Phi\rangle}{\Lambda} = \frac{1}{B},
\end{equation}
where in this work we take
\begin{equation}
  B = \frac{75}{14} \simeq 5.357,
\end{equation}
and write all hierarchies in terms of powers of $B$ (or equivalently $\epsilon$).

\begin{figure*}[!tbp]
\centering
\begin{tikzpicture}[
  x=1.0cm,y=0.9cm,
  >=Latex,
  ferm/.style={-Latex,thick},
  heavy/.style={double, line width=0.6pt, double distance=1.1pt},
  scal/.style={-Latex,thick,densely dashed},
  vtx/.style={circle,draw,fill=black,inner sep=1.2pt},
  lab/.style={font=\scriptsize}
]
\node[lab] at (0,1.55) {$Q_i$};
\node[lab] at (8,1.55) {$f^c_j$};
\node[lab] (H)  at (4,-2.0) {$H$};

\foreach \idx/\yy/\ph/\pow/\dy/\dph in {
  1/1.2/{$1$}/{p^f_{ij}}/0/0,
  2/0.0/{$e^{i\phi_f}$}/{p^f_{ij}+\Delta^f_{ij}}/-0.25/0.15,
  3/-1.2/{$e^{i\psi_f}$}/{p^f_{ij}+\Delta^{\prime f}_{ij}}/-0.25/0.15
}{
  \node[vtx] (L\idx) at (1,\yy) {};
  \node[vtx] (A\idx) at (3,\yy) {};
  \node[vtx] (B\idx) at (5,\yy) {};
  \node[vtx] (R\idx) at (7,\yy) {};

  \draw[ferm] (0,\yy) -- (L\idx);
  \draw[ferm] (R\idx) -- (8,\yy);

  \draw[heavy] (L\idx) -- (A\idx);
  \draw[heavy] (A\idx) -- (B\idx);
  \draw[heavy] (B\idx) -- (R\idx);

  \draw[scal] (H) -- (A\idx);

  \draw[scal] (B\idx) -- ++(0.0,0.65);
  \node[lab,anchor=south,yshift=\dy cm] at ($(B\idx)+(0,0.65)$) {$\Phi^{\,\pow}$};

  \node[lab,anchor=east,yshift=\dph cm] at (8.0,\yy) {\ph};
}

\node[lab] at (4,1.55) {\textit{chain 1}};
\node[lab] at (4,0.35) {\textit{chain 2}};
\node[lab] at (4,-0.85) {\textit{chain 3}};
\end{tikzpicture}
\caption{Schematic three-messenger realization of a single Yukawa entry
$(Y_f)_{ij}$ in the FN framework~\cite{Froggatt:1978nt,Leurer:1992qj,Leurer:1993gy}.  The external legs are the left-handed quark
doublet $Q_i$, the right-handed singlet $f^c_j$ ($f=u,d$), and the Higgs
doublet $H$. Heavy vector-like messenger fermions (doublets and singlets) run in the internal lines. Each messenger chain carries a certain number of flavon insertions $\Phi$, generating a power of $\epsilon = \langle \Phi\rangle/\Lambda = 1/B$.
Summing the three chains yields an effective suppression $\epsilon^{p^f_{ij}}$
times an $\ordone$ complex coefficient $C^f_{ij}$.}
\label{fig:messenger-feynman}
\end{figure*}
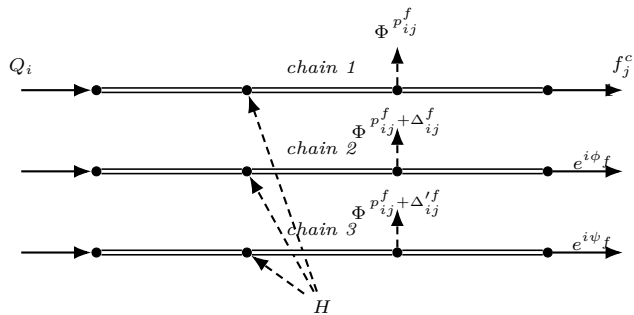

We work in a one-flavon FN framework in which each Yukawa entry arises from a
small number of messenger chains.  At the effective level this means
\begin{equation}
  (Y_f)_{ij} \sim C^f_{ij}\,\epsilon^{p^f_{ij}},
  \qquad f=u,d,e,\nu,
\end{equation}
with $C^f_{ij}=\ordone$ complex coefficients and rational exponents
$p^f_{ij}$ fixed by the underlying FN charges.  Our particular UV
interpretation is that each $(Y_f)_{ij}$ is generated by three messenger chains
with unit-magnitude couplings (some of which carry an overall minus sign)
and the effective $\ordone$ coefficients $C^f_{ij}$ arise from the
coherent sum of these chains.  A schematic version of this messenger picture
is shown in Fig.~\ref{fig:messenger-feynman}.  Related one-flavon implementations and additional phenomenological details are discussed in Ref.~\cite{Barger:2025bfn}.

\vspace{0.3em}
\noindent\textbf{Motivation for the two-over-two lattice.}
A central empirical observation in this paper is that certain dimensionless
combinations of fermion masses built as ``two-over-two'' (quadrilateral) ratios
\begin{equation}
  R_{abcd} \equiv \frac{m_a\,m_d}{m_b\,m_c}
\end{equation}
cluster numerically very close to integer powers of the same parameter $B$,
\begin{equation}
  R_{abcd} \simeq B^{\,n_{abcd}}  
\end{equation}
when evaluated with running quark masses at the electroweak scale.
Examples include ratios such as
$(m_s^2)/(m_u m_b)$, $(m_u m_t)/(m_c m_s)$, $(m_c m_t)/(m_d m_s)$, and others.
Taken together, these relations suggest that the quark masses populate a
lattice in $\log_B m$ space, with a small set of integer exponents
controlling many different mass combinations.

To visualize this structure, it is convenient to assign integer
coordinates $(x,y)$ to each ratio by
\begin{equation}
  y\equiv 9\,(n_a+n_d), \qquad
  x \equiv 9\,(n_b+n_c),
\end{equation}
so that
\begin{equation}
  n_{abcd} = \frac{y - x}{9}.
\end{equation}
Each two-over-two relation then appears as a point in the $(x,y)$ plane,
and all ratios with the same power $n_{abcd}$ lie on a straight line
$y-x = 9\,n_{abcd}$.  We refer to this picture as the
two-over-two lattice.

\vspace{0.3em}
\noindent\textbf{Goals and scope of this work.}
The goals of this paper are:
\begin{itemize}
  \item to show that a single parameter $B \simeq 5.357$, together with rational
        FN exponents $p^f_{ij}$ on a fine lattice, provides a good description
        of the observed quark and charged-lepton mass hierarchies at $M_Z$;
  \item to connect these exponent textures to a simple three-messenger FN
        picture with unit-magnitude UV couplings, as sketched in
        Fig.~\ref{fig:messenger-feynman};
  \item to relate the exponent textures to the two-over-two lattice in
        $(x,y)$ space and to a compact set of FN charges for quarks and
        leptons; and
  \item to outline how neutrino mass hierarchies can be embedded into the same
        $B^n$ counting, leaving a detailed PMNS analysis to future work.
\end{itemize}

\noindent\textbf{Notation and conventions.}
Throughout the paper we use:
\begin{itemize}
  \item $B \simeq 5.357$ and $\epsilon \equiv 1/B$ as the fundamental hierarchy
        parameter;
  \item running quark and charged-lepton masses in the $\MS$ scheme evolved to
        the scale $M_Z$~\cite{Huang:2021rkm,PDG:2024isi};
  \item $p^f_{ij}$ for FN exponents in the Yukawa matrices $Y_f$
        ($f=u,d,e$), with $(Y_f)_{ij} \sim C^f_{ij} \epsilon^{p^f_{ij}}$ and
        $C^f_{ij} = \ordone$;
  \item $R_{abcd} = (m_a m_d)/(m_b m_c)$ for two-over-two mass ratios; and
  \item $(x,y)$ for the integer coordinates associated with a given
        two-over-two ratio, defined above.
\end{itemize}

The remainder of the paper is organized as follows.
In Sec.~\ref{sec:fncharges} we present the rational FN exponent textures for the quark Yukawa
matrices and the corresponding FN charge assignments for quarks and leptons.
In Sec.~\ref{sec:chargedleptons} we analyze the charged-lepton hierarchy in the same $B^n$
language.  Section~\ref{sec:results} discusses the numerical quality of the quark mass fit
and the structure of the two-over-two lattice in $(x,y)$ space.
Neutrino mass hierarchies within the $B^n$ framework are introduced in Sec.~\ref{sec:neutrino}.
A discrete $Z_N$ interpretation of the rational FN charges is outlined in Sec.~\ref{sec:ZN}.
We summarize our main conclusions and comment on future directions in Sec.~\ref{sec:outlook}.
Several technical details and explicit benchmark parameter choices are collected
in Appendix~\ref{app:3mess-params}, and additional two-over-two lattice details are given in Appendix~\ref{app:kspace}.

\begin{table}[!tbp]
\caption{One-flavon FN charges for quarks and leptons. Quark charges are extracted from the exponent matrices in Eq.~\eqref{eq:Y-diag-def} using
$Q(Q_3)=Q(u^c_3)=Q(d^c_3)=0$ and $Q(H)=0$.
Lepton-doublet charges follow from the Weinberg-operator exponents in Eq.~\eqref{eq:mnu-p-matrix} via Eq.~\eqref{eq:pnu-factor} with $Q(L_3)=0$,
and charged-lepton singlet charges are fixed from the diagonal charged-lepton scaling (Sec.~\ref{sec:chargedleptons}) assuming diagonal dominance of $Y_e$.
The rightmost column lists charges in integer units of $1/18$.}
\renewcommand{\arraystretch}{1.5}
\begin{ruledtabular}
\begin{tabular}{lcc}
Field & $Q(\cdot)$ & $18\,Q(\cdot)$ \\
\hline
\multicolumn{3}{c}{Quarks}\\
\hline
$Q_1$ & $3$ & $54$ \\
$Q_2$ & $2$ & $36$ \\
$Q_3$ & $0$ & $0$ \\
\hline
$u^c_1$ & $37/9$ & $74$ \\
$u^c_2$ & $4/3$ & $24$ \\
$u^c_3$ & $0$ & $0$ \\
\hline
$d^c_1$ & $10/9$ & $20$ \\
$d^c_2$ & $1/3$ & $6$ \\
$d^c_3$ & $0$ & $0$ \\
\hline\hline
\multicolumn{3}{c}{Leptons (Weinberg operator)}\\
\hline
$L_1$ & $1$ & $18$ \\
$L_2$ & $1/2$ & $9$ \\
$L_3$ & $0$ & $0$ \\
\hline
$e^c_1$ & $23/6$ & $69$ \\
$e^c_2$ & $7/6$ & $21$ \\
$e^c_3$ & $0$ & $0$ \\
\end{tabular}
\end{ruledtabular}
\renewcommand{\arraystretch}{1.0}
\label{tab:fncharges}
\end{table}

\section{Quark Yukawa textures and FN exponents}
\label{sec:fncharges}

We parameterize the effective quark Yukawas as
\begin{equation}
\begin{aligned}
  (Y_u)_{ij} &\sim C^{u}_{ij}\,\e^{\,p^{u}_{ij}},\\
  (Y_d)_{ij} &\sim C^{d}_{ij}\,\e^{\,p^{d}_{ij}},
\end{aligned}
\end{equation}
with exponent matrices
\begin{equation}
\begin{aligned}
  p^{u}_{ij} &=
  {\setlength{\arraycolsep}{6pt}\renewcommand{\arraystretch}{1.15}
  \begin{pmatrix}
    \tfrac{64}{9} & \tfrac{13}{3} & 3\\
    \tfrac{55}{9} & \tfrac{10}{3} & 2\\
    \tfrac{37}{9} & \tfrac{4}{3}  & 0
  \end{pmatrix}},\\[8pt]
  p^{d}_{ij} &=
  {\setlength{\arraycolsep}{6pt}\renewcommand{\arraystretch}{1.15}
  \begin{pmatrix}
    \tfrac{37}{9} & \tfrac{10}{3} & 3\\
    \tfrac{28}{9} & \tfrac{7}{3}  & 2\\
    \tfrac{10}{9} & \tfrac{1}{3}  & 0
  \end{pmatrix}}.
\end{aligned}
\label{eq:Y-diag-def}
\end{equation}

Eq.~\eqref{eq:Y-diag-def} is motivated by minimality and by the
quark mass spectrum alone. With a single flavon and
$Q(H) = 0$, FN predicts that Yukawa suppressions factorize into additive left- and right-handed charges, 
implying $p^u_{ij}=Q(Q_i)+Q(u^c_j)$ and $p^d_{ij}=Q(Q_i)+Q(d^c_j)$.

Independently, the observed quark masses
satisfy a set of near-integer ``two-over-two'' relations
$(m_a m_d)/(m_b m_c)\simeq B^n$, showing that the spectrum lies
close to a low-dimensional lattice in $\log_B m$. Eq.~\eqref{eq:Y-diag-def}
is the minimal matrix-level completion compatible with
these mass-lattice regularities; CKM phenomenology
then probes the off-diagonal structure and the size of the
effective $\ordone$ factors and is deferred to separate work.

In the multi-messenger representation discussed below, all UV magnitudes are taken to be unity (some with an overall sign $-1$) and the only free parameters at this stage are relative phases.
For $\e\ll1$ each entry is dominated by the smallest exponent in the sum; this is the familiar sequential-dominance limit~\cite{Froggatt:1978nt}.

We measure FN exponents in units of $1/9$, so that an integer power $n_{ij}$ of $(\Phi/\Lambda)$ corresponds to a rational lattice exponent $p^f_{ij} = n_{ij}/9$ in Eq.~\eqref{eq:Y-diag-def}.

\subsection{Sequential dominance from three messenger chains}
\label{sec:seqdom}

Each effective Yukawa element is interpreted as the sum over a small number of heavy-messenger chains,
\begin{equation}
  (Y_f)_{ij} = \sum_{a=1}^{N_f} e^{i\phi^f_a}\,\e^{\,p_{ij}^{f(a)}},
  \qquad f=u,d,
  \label{eq:multi-messenger}
\end{equation}
with all UV magnitudes set to unity.  For concreteness we take a minimal three-chain ansatz,
\begin{equation}
\begin{aligned}
  Y_u &= \e^{\,p^u}+e^{i\phi_u}\,\e^{\,p^u+\Delta_u}+e^{i\psi_u}\,\e^{\,p^u+\Delta'_u},\\
  Y_d &= \e^{\,p^d}+e^{i\phi_d}\,\e^{\,p^d+\Delta_d}+e^{i\psi_d}\,\e^{\,p^d+\Delta'_d},
\end{aligned}
\label{eq:three-term}
\end{equation}
where $\Delta,\Delta'\ge 0$ are entry-dependent additional suppression matrices
(with the same rational lattice spacing as $p^{u,d}$).  Each element has the form
\begin{equation}
  (Y_f)_{ij}=\e^{\,p^f_{ij}}\Big[1+e^{i\phi_f}\e^{\Delta^f_{ij}}+e^{i\psi_f}\e^{\Delta^{\prime f}_{ij}}\Big],
  \qquad f=u,d,
  \label{eq:Ceffective}
\end{equation}
so the effective coefficients
\begin{equation}
  C^{f}_{ij}=1+e^{i\phi_f}\e^{\Delta^f_{ij}}+e^{i\psi_f}\e^{\Delta^{\prime f}_{ij}}
\end{equation}
are automatically $\ordone$ and can carry nontrivial phases.
This three-chain realization provides a concrete UV-friendly origin for the emergent $\ordone$ coefficients that appear in many FN-based flavor analyses~\cite{Leurer:1992qj,Leurer:1993gy,Barger:2025bfn}.  Explicit benchmark shifts and phases are given in Appendix~\ref{app:3mess-params}.

\subsection{FN charge assignments}
\label{sec:charges}

Assuming a single flavon and a Higgs neutral under the flavor symmetry ($Q(H)=0$), the FN exponents factorize as~\cite{Froggatt:1978nt,Leurer:1992qj,Leurer:1993gy}
\begin{equation}
p^u_{ij}=Q(Q_i)+Q(u^c_j),\qquad
p^d_{ij}=Q(Q_i)+Q(d^c_j),
\label{eq:pij-factor}
\end{equation}
up to an overall shift of the charge origin.
A convenient convention is $Q(Q_3)=Q(u^c_3)=Q(d^c_3)=0$, consistent with $p^{u}_{33}=p^{d}_{33}=0$.
Then
\begin{equation}
Q(Q_i)=p^u_{i3}=p^d_{i3},\qquad
Q(u^c_j)=p^u_{3j},\qquad
Q(d^c_j)=p^d_{3j}.
\label{eq:charges-read}
\end{equation}

For leptons we commit here to the Weinberg-operator description of light neutrino masses~\cite{Weinberg:1979sa}, so that
\begin{equation}
p^\nu_{ij}=Q(L_i)+Q(L_j),
\label{eq:pnu-factor}
\end{equation}
with $p^\nu_{ij}$ specified below in Eq.~\eqref{eq:mnu-p-matrix} and $Q(L_3)=0$.
Charged-lepton charges then follow from $p^e_{ij}=Q(L_i)+Q(e^c_j)$ and the diagonal mass scaling.

The resulting one-flavon FN charges are summarized in Table~\ref{tab:fncharges}.

\begin{table}[!tbp]
\caption{Charged-lepton $\MS$ running masses at $M_Z$ are taken from Huang and Zhou~\cite{Huang:2021rkm}. Pole masses are from the PDG review~\cite{PDG:2024isi}.}
\renewcommand{\arraystretch}{1.5}
\begin{ruledtabular}
\begin{tabular}{lcc}
Observable & Pole [MeV] & $m_\ell^{\MS}(M_Z)$ [MeV] \\
\hline
$m_e$   & 0.51099895  & 0.48307 \\
$m_\mu$ & 105.6583755 & 101.766 \\
$m_\tau$& 1776.93     & 1728.56 \\
\end{tabular}
\end{ruledtabular}
\renewcommand{\arraystretch}{1.0}
\label{tab:leptonmasses}
\end{table}

\section{Charged leptons: \texorpdfstring{$B$}{B}-power relations}
\label{sec:chargedleptons}

Using the $\MS$  running masses from Table~\ref{tab:leptonmasses}, the charged-lepton hierarchy is well captured by
\begin{equation}
  \Bpar^3 \simeq \left(\frac{m_\mu^2}{m_e\,m_\tau}\right)^{\!2},
  \qquad
  \Bpar^8 \simeq \frac{m_\mu\,m_\tau}{m_e^{\,2}}.
  \label{eq:B-relations-leptons}
\end{equation}
Writing $m_\alpha \propto \Bpar^{n_\alpha}$ and fixing $n_\tau=0$ gives
\begin{align}
  3 &= 4n_\mu-2n_e, \label{eq:lepton-n-eq1}\\
  8 &= n_\mu-2n_e. \label{eq:lepton-n-eq2}
\end{align}
Solving,
\begin{equation}
  n_e=-\frac{29}{6},\qquad n_\mu=-\frac{5}{3},\qquad n_\tau=0.
  \label{eq:lepton-n-sol}
\end{equation}
Equivalently, in FN form $m\propto \e^{\,p}$ with $p\equiv -n$,
\begin{equation}
  m_e:m_\mu:m_\tau \propto \e^{29/6}:\e^{5/3}:1,
  \label{eq:lepton-p-sol}
\end{equation}
consistent with the lepton charges quoted earlier.

\section{Two-over-two lattice and quark masses}
\label{sec:results}

Representative two-over-two relations are listed in Table~\ref{tab:twoover}.

\begin{table}[!tbp]
\caption{Two-over-two quark mass ratios and corresponding integer $B$-powers.  Here $u,d,s,c,b,t$ denote running quark masses in the $\MS$ scheme at $M_Z$.  When multiple ratios share the same $n$, they are displayed as equalities.}
\renewcommand{\arraystretch}{1.5}
\begin{ruledtabular}
\begin{tabular}{lc}
$B^{\,n}$ & Representative ratio(s) $R_n$ \\
\hline
$B^{0}$ & $(ss)/(ub)$ \\
$B^{1}$ & $(ut)/(cs) = (st)/(cb)$ \\
$B^{2}$ & $(dc)/(su)$ \\
$B^{3}$ & $(dt)/(bu)$ \\
$B^{5}$ & $(bb)/(cd) = (sc)/(dd)$ \\
$B^{6}$ & $(ut)/(dd)$ \\
$B^{7}$ & $(bb)/(su) = (sb)/(uu) = (cc)/(ud)$ \\
$B^{8}$ & $(ct)/(ds)$ \\
$B^{9}$ & $(tt)/(db)$ \\
\end{tabular}
\end{ruledtabular}
\renewcommand{\arraystretch}{1.0}
\label{tab:twoover}
\end{table}

We introduce $B$-exponents $n_f$ for each quark flavor $f$ by
\begin{equation}
  m_f = m_b^{\rm ref}\,\Bpar^{\,n_f},\qquad n_b=0,
\end{equation}
and take as a reference the golden ratio for the b-mass to $\tau$-mass ratio:
\begin{equation}
  m_b^{\rm ref} \equiv \phiGR\,m_\tau^{\MS}(M_Z),
  \qquad
  \phiGR \equiv \frac{1+\sqrt{5}}{2}.
  \label{eq:mbref}
\end{equation}

A convenient independent set of constraints from Table~\ref{tab:twoover} is
\begin{align}
  0&= 2n_s-n_u, &
  1&= n_s+n_t-n_c, \nonumber\\[-3pt]
  2&= n_d+n_c-n_s-n_u, &
  5&= -n_c-n_d, \nonumber\\[1pt]
  9&= 2n_t-n_d. &&
  \label{eq:nf-system}
\end{align}

Solving,
\begin{equation}
\begin{aligned}
  n_{d} &=-\tfrac{37}{9},\quad
  n_{s}=-\tfrac{7}{3},\quad
  n_{b}=0,\\
  n_{u} &=-\tfrac{14}{3},\quad
  n_{c}=-\tfrac{8}{9},\quad
  n_{t}= \tfrac{22}{9}.
\end{aligned}
\label{eq:nf-solution}
\end{equation}

The running quark masses used for numerical benchmarks are summarized in Table~\ref{tab:quarkmasses}.

\begin{table}[!tbp]
\caption{Running quark masses at $M_Z$ used for the numerical examples. Central values follow recent electroweak-scale determinations and PDG compilations~\cite{Huang:2021rkm,PDG:2024isi}; the ``Model'' column gives the adopted benchmark values.}
\renewcommand{\arraystretch}{1.5}
\begin{ruledtabular}
\begin{tabular}{lcc}
Observable & PDG$\to M_Z$ & Model \\
\hline
$m_u$ [MeV] & $1.11 \pm 0.20$   & 1.11 \\
$m_d$ [MeV] & $2.82 \pm 0.20$   & 2.82 \\
$m_s$ [MeV] & $55.7 \pm 1.3$    & 55.7 \\
$m_c$ [GeV] & $0.629 \pm 0.012$ & 0.629 \\
$m_b$ [GeV] & $2.794 \pm 0.034$ & 2.794 \\
$m_t$ [GeV] & $169 \pm 1.1$     & 169 \\
\end{tabular}
\end{ruledtabular}
\renewcommand{\arraystretch}{1.0}
\label{tab:quarkmasses}
\end{table}

To visualize the lattice, write
\begin{equation}
R_{abcd}=\frac{m_a m_d}{m_b m_c}
\end{equation}
\begin{equation}
\qquad
\log_{\Bpar} R_{abcd} = (n_a+n_d)-(n_b+n_c)\equiv n_{abcd}.
\end{equation}
Define integer coordinates
\begin{equation}
  y \equiv 9\,(n_a+n_d),\qquad
  x \equiv 9\,(n_b+n_c),
\end{equation}
so that
\begin{equation}
  n_{abcd}=\frac{y-x}{9}.
  \label{eq:n-from-k}
\end{equation}

Representative coordinates are listed in Table~\ref{tab:kXY} and plotted in Fig.~\ref{fig:k6k7}; further lattice details are given in Appendix~\ref{app:kspace}.

\begin{table}[!tbp]
\caption{Representative two-over-two ratios $R_{abcd}$ and their integer $(x,y)$ coordinates, computed from the exponents in Eq.~\eqref{eq:nf-solution}.  The $B$-power is $n_{abcd}=(y-x)/9$ [Eq.~\eqref{eq:n-from-k}].}
\renewcommand{\arraystretch}{1.5}
\begin{ruledtabular}
\begin{tabular}{lccc}
Ratio $R_{abcd}$ & $n_{abcd}$ & $x$ & $y$ \\
\hline
$(ss)/(ub)$ & $0$ & $-42$ & $-42$ \\
$(ut)/(cs)$ & $1$ & $-29$ & $-20$ \\
$(st)/(cb)$ & $1$ & $-8$  & $1$   \\
$(dc)/(su)$ & $2$ & $-63$ & $-45$ \\
$(dt)/(bu)$ & $3$ & $-42$ & $-15$ \\
$(bb)/(cd)$ & $5$ & $-45$ & $0$   \\
$(sc)/(dd)$ & $5$ & $-74$ & $-29$ \\
$(ut)/(dd)$ & $6$ & $-74$ & $-20$ \\
$(bb)/(su)$ & $7$ & $-63$ & $0$   \\
$(sb)/(uu)$ & $7$ & $-84$ & $-21$ \\
$(cc)/(ud)$ & $7$ & $-79$ & $-16$ \\
$(ct)/(ds)$ & $8$ & $-58$ & $14$  \\
$(tt)/(db)$ & $9$ & $-37$ & $44$  \\
\end{tabular}
\end{ruledtabular}
\renewcommand{\arraystretch}{1.0}
\label{tab:kXY}
\end{table}

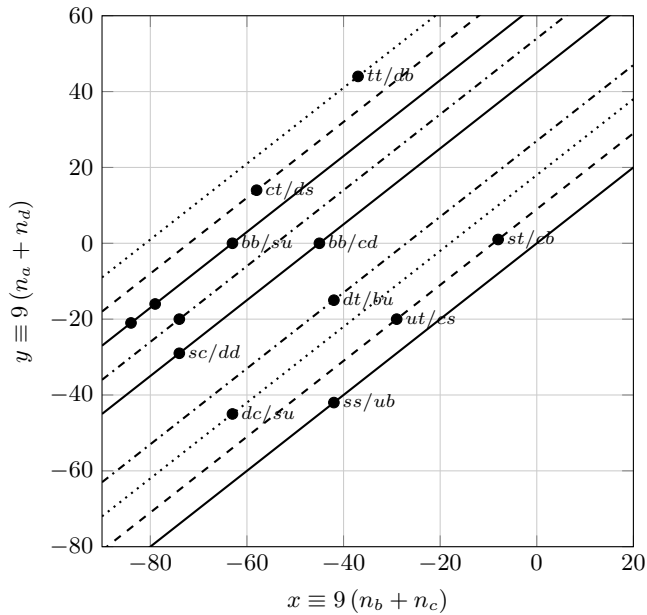
\begin{figure*}[!tbp]
\centering
\begin{tikzpicture}
\begin{axis}[
  width=1\columnwidth,
  height=1\columnwidth,
  xlabel={$x\equiv 9\,(n_b+n_c)$},
  ylabel={$y\equiv 9\,(n_a+n_d)$},
  xmin=-90, xmax=20,
  ymin=-80, ymax=60,
  grid=both,
  major grid style={line width=0.4pt,draw=gray!40},
  minor grid style={line width=0.2pt,draw=gray!20},
  ticklabel style={font=\small},
  label style={font=\small},
]
\addplot[domain=-90:20, samples=2, thick] {x};              
\addplot[domain=-90:20, samples=2, thick, dashed] {x+9};    
\addplot[domain=-90:20, samples=2, thick, dotted] {x+18};   
\addplot[domain=-90:20, samples=2, thick, dashdotted] {x+27}; 
\addplot[domain=-90:20, samples=2, thick] {x+45};           
\addplot[domain=-90:20, samples=2, thick, dashdotted] {x+54}; 
\addplot[domain=-90:20, samples=2, thick] {x+63};           
\addplot[domain=-90:20, samples=2, thick, dashed] {x+72};   
\addplot[domain=-90:20, samples=2, thick, dotted] {x+81};   

\addplot[only marks, mark=*] coordinates {
  (-42,-42) (-29,-20) (-8,1) (-63,-45) (-42,-15) (-45,0)
  (-74,-29) (-74,-20) (-63,0) (-84,-21) (-79,-16) (-58,14) (-37,44)
};

\node[anchor=west, font=\scriptsize] at (axis cs:-42,-42) {$ss/ub$};
\node[anchor=west, font=\scriptsize] at (axis cs:-29,-20) {$ut/cs$};
\node[anchor=west, font=\scriptsize] at (axis cs:-8,1) {$st/cb$};
\node[anchor=west, font=\scriptsize] at (axis cs:-63,-45) {$dc/su$};
\node[anchor=west, font=\scriptsize] at (axis cs:-42,-15) {$dt/bu$};
\node[anchor=west, font=\scriptsize] at (axis cs:-45,0) {$bb/cd$};
\node[anchor=west, font=\scriptsize] at (axis cs:-74,-29) {$sc/dd$};
\node[anchor=west, font=\scriptsize] at (axis cs:-63,0) {$bb/su$};
\node[anchor=west, font=\scriptsize] at (axis cs:-58,14) {$ct/ds$};
\node[anchor=west, font=\scriptsize] at (axis cs:-37,44) {$tt/db$};
\end{axis}
\end{tikzpicture}
\caption{Two-over-two relations as points in the integer $(x,y)$ plane.
Each ratio $R_{abcd}$ lies on a line of constant $n_{abcd}$ given by Eq.~\eqref{eq:n-from-k}.
The straight lines (from bottom to top at fixed $X$) correspond to $n=0,1,2,3,5,6,7,8,9$.
All lattice points are plotted; the $(x,y)$ coordinates are listed in Table~\ref{tab:kXY}.  The horizontal axis is $x\equiv 9\,(n_b+n_c)$ and the vertical axis is $y\equiv 9\,(n_a+n_d)$. The axes are linear in the integers $(x,y)$ (i.e. not log--scaled), but they encode logarithmic ratio information: if $R_{abcd}\sim B^{-n_{abcd}}$ with $n_{abcd}=(y-x)/9$, then $n_{abcd}=-\log_B R_{abcd}$.}
\label{fig:k6k7}

\end{figure*}

\subsection{Quark flavor points in the \texorpdfstring{$(x^{\prime},y^{\prime})$}{(x',y')} plane}

For individual quark flavors it is useful to plot the integer FN charges (in units of $1/9$) as a two-dimensional point.
Using the conventions of Sec.~\ref{sec:fncharges}, we define
\begin{equation}
  y^{\prime (f)} \equiv 9\,[Q(Q_i)-Q(Q_3)],\qquad
  x^{\prime (f)} \equiv 9\,[Q(f^c_i)-Q(f^c_3)],
\label{eq:k6k7-quarks}
\end{equation}
where $f^c_j=u^c_j$ for $f=u,c,t$ and $f^c_j=d^c_j$ for $f=d,s,b$.
In the diagonal-dominant limit, $p^{u}_{ii}=(y^{\prime (u_i)}+x^{\prime (u_i)})/9$ and $p^{d}_{ii}=(y^{\prime (d_i)}+x^{\prime (d_i)})/9$.

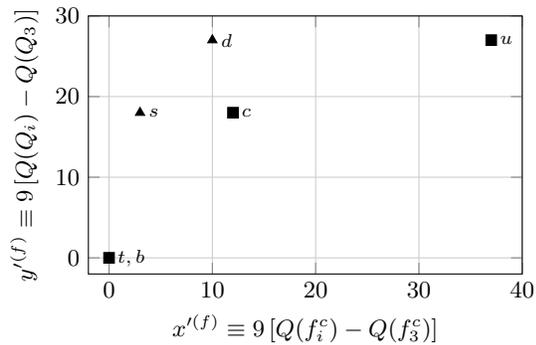
\begin{figure}[t]
\centering
\begin{tikzpicture}
\begin{axis}[
  width=0.85\columnwidth,
  height=0.58\columnwidth,
  xlabel={$x^{\prime (f)} \equiv 9\,[Q(f^c_i)-Q(f^c_3)]$},
  ylabel={$y^{\prime (f)} \equiv 9\,[Q(Q_i)-Q(Q_3)]$},
  xmin=-2, xmax=40,
  ymin=-2, ymax=30,
  grid=both,
  major grid style={line width=0.4pt,draw=gray!40},
  minor grid style={line width=0.2pt,draw=gray!20},
  ticklabel style={font=\small},
  label style={font=\small},
]
\addplot[only marks, mark=square*] coordinates {(37,27) (12,18) (0,0)};
\addplot[only marks, mark=triangle*] coordinates {(10,27) (3,18) (0,0)};

\node[anchor=west, font=\scriptsize] at (axis cs:37,27) {$u$};
\node[anchor=west, font=\scriptsize] at (axis cs:12,18) {$c$};
\node[anchor=west, font=\scriptsize] at (axis cs:10,27) {$d$};
\node[anchor=west, font=\scriptsize] at (axis cs:3,18) {$s$};
\node[anchor=west, font=\scriptsize] at (axis cs:0,0) {$t,b$};
\end{axis}
\end{tikzpicture}
\caption{Quark flavor points in the $(x^{\prime},y^{\prime})$ plane constructed from the FN charges in Table~\ref{tab:fncharges} via Eq.~\eqref{eq:k6k7-quarks}. Up-type quarks are shown as squares and down-type quarks as triangles. The origin is chosen so that the third generation $(t,b)$ sits at $(0,0)$, which is invariant under uniform shifts of all charges.}
\label{fig:k6k7-quarks}
\end{figure}

Figure~\ref{fig:k6k7-quarks} 
shows the resulting quark points, constructed from the charges in Table~\ref{tab:fncharges}.

\section{Neutrino masses in the \texorpdfstring{$\Bpar^n$}{B\textsuperscript{n}} framework}
\label{sec:neutrino}

We treat neutrino masses at the level of eigenvalues, postponing PMNS mixing and CP phases to a companion paper.

We parameterize the effective light-neutrino Majorana mass matrix as
\begin{equation}
\begin{aligned}
  (m_\nu)_{ij} &\sim m_0\,C^{\nu}_{ij}\,\e^{\,p^{\nu}_{ij}},\\[4pt]
  p^{\nu}_{ij} &=
  \begin{pmatrix}
    2 & \tfrac{3}{2} & 1\\
    \tfrac{3}{2} & 1 & \tfrac{1}{2}\\
    1 & \tfrac{1}{2} & 0
  \end{pmatrix},
\end{aligned}
\label{eq:mnu-p-matrix}
\end{equation}
with $C^\nu_{ij}=\ordone$.
This corresponds, for example, to lepton-doublet charges $Q(L_i)=(1,\tfrac12,0)$ in a one-flavon Weinberg-operator realization~\cite{Weinberg:1979sa}.

For normal ordering, neutrino-oscillation data determine two independent mass-squared splittings; representative global-fit values are~\cite{NuFit:2024,Capozzi:2025}
\begin{equation}
\Delta m_{21}^2 \simeq 7.4\times 10^{-5}\ \mathrm{eV}^2,
\qquad
\Delta m_{31}^2 \simeq 2.5\times 10^{-3}\ \mathrm{eV}^2.
\end{equation}
Motivated by the $B$-tower structure, we adopt
\begin{equation}
m_3 = m_0,\qquad
m_2 = \frac{c_2\,m_0}{\Bpar},\qquad
m_1 = \frac{c_1\,m_0}{\Bpar^2},
\end{equation}
with $c_{1,2}=\ordone$.

For $c_1=c_2=1$, the atmospheric splitting fixes
\begin{equation}
m_0 = \sqrt{\frac{\Delta m_{31}^2}{1-\Bpar^{-4}}}
\simeq 5.0\times 10^{-2}\ \mathrm{eV},
\end{equation}
so that
\begin{equation}
(m_3,m_2,m_1) \simeq (5.0,\,0.93,\,0.17)\times 10^{-2}\ \mathrm{eV}.
\end{equation}
The implied solar splitting is
\begin{equation}
\Delta m_{21}^2 = m_2^2-m_1^2
\simeq 8\times 10^{-5}\ \mathrm{eV}^2,
\end{equation}
already close to the empirical value within $\ordone$ factors.
The sum of masses is
\begin{equation}
\sum_i m_i \simeq m_0\left(1+\frac{1}{\Bpar}+\frac{1}{\Bpar^2}\right)
\simeq 6.1\times 10^{-2}\ \mathrm{eV},
\end{equation}
compatible with current cosmological bounds.

\section{Discrete \texorpdfstring{$Z_N$}{ZN} interpretation}
\label{sec:ZN}

The rational charges in Table~\ref{tab:fncharges} can be converted to integers by multiplying by a common denominator (here $18$) and working modulo $N$:
\begin{equation}
  q_f \equiv 18\,Q(f) \pmod N.
\end{equation}
Yukawa suppressions then depend only on these discrete charges and the number of flavon insertions required to form an invariant operator.
A convenient choice is $N=18$, which accommodates the $1/18$ lattice while allowing a nontrivial discrete symmetry.

For a discrete gauge symmetry, consistency requires that the mixed gauge--$Z_N$ anomalies satisfy the discrete anomaly cancellation conditions, e.g.\ the Ibáñez--Ross constraints~\cite{Krauss:1988zc,BanksDine:1991xc,Ibanez:1991hc}.
We do not specify a UV completion here; the $Z_N$ viewpoint is used as a bookkeeping device and a guide to possible anomaly-free completions.

\section{Summary and outlook}
\label{sec:outlook}

We have presented a one-flavon FN framework in which the dominant hierarchies of quark and lepton Yukawa couplings (and the light-neutrino Majorana mass matrix via the Weinberg operator) are controlled by powers of a single parameter $\e=1/\Bpar$, with a small rational lattice of exponents capturing the remaining structure.

Empirically, many quadrilateral mass ratios of the form $(m_a m_d)/(m_b m_c)$ cluster near integer powers of $\Bpar$, and this two-over-two lattice structure lifts naturally to matrix-level FN exponent textures and compact charge assignments.  A minimal three-messenger construction with unit-magnitude UV couplings realizes the required hierarchy while leaving only $\ordone$ complex factors in the low-energy Yukawa couplings.

The structure of the lattice is visualized in the $(x,y)$ plane in Fig.~\ref{fig:k6k7} and in the $(x',y')$ plane in Fig.~\ref{fig:k6k7-quarks}, and the underlying charge and mass inputs are summarized in Tables~\ref{tab:fncharges}, \ref{tab:twoover}, \ref{tab:quarkmasses} and~\ref{tab:kXY}.
A companion paper will present the detailed CKM and PMNS phenomenology (including Dirac CP phases) in this three-messenger, two-over-two lattice framework.

\begin{acknowledgments}
VB gratefully acknowledges support from the William F. Vilas Estate.
\end{acknowledgments}

\appendix

\section{Benchmark three-messenger input}
\label{app:3mess-params}

We use
\begin{equation}
  \Bpar = 5.357,\qquad \e=\frac{1}{\Bpar}\,,
\end{equation}
and the exponent matrices $p^{u}$ and $p^{d}$ in Eq.~\eqref{eq:Y-diag-def}.

For each sector $f=u,d$ we take factorized nonnegative shifts
\begin{equation}
  \Delta^{f}_{ij}=\alpha^{f}_i+\beta^{f}_j,\qquad
  \Delta^{\prime f}_{ij}=\alpha^{\prime f}_i+\beta^{\prime f}_j,
\end{equation}
with
\begin{equation}
\begin{aligned}
\alpha^{u}&=\tfrac{1}{9}(0,2,1),&
\beta^{u}&=\tfrac{1}{9}(0,3,3),\\
\alpha^{\prime u}&=\tfrac{1}{9}(2,1,0),&
\beta^{\prime u}&=\tfrac{1}{9}(0,1,0),\\[3pt]
\alpha^{d}&=\tfrac{1}{9}(5,0,5),&
\beta^{d}&=\tfrac{1}{9}(0,3,4),\\
\alpha^{\prime d}&=\tfrac{1}{9}(2,1,0),&
\beta^{\prime d}&=\tfrac{1}{9}(0,1,0).
\end{aligned}
\label{eq:app-shifts}
\end{equation}
\noindent
These benchmark shifts are chosen so that, with unit-modulus messenger-chain coefficients and the phases in Eq.~\eqref{eq:app-phases}, the singular values of $Y_u$ and $Y_d$ (after overall normalization to $m_t$ and $m_b$) reproduce the quark masses in Table~\ref{tab:quarkmasses}.

The effective $\ordone$ factors are
$C^{f}_{ij}=1+e^{i\phi_f}\e^{\Delta^{f}_{ij}}+e^{i\psi_f}\e^{\Delta^{\prime f}_{ij}}$.
A benchmark choice of phases (in radians) is
\begin{equation}
  \phi_u\simeq 0,\quad \psi_u\simeq 0,\qquad 
  \phi_d\simeq 2.8596,\quad \psi_d\simeq 3.540.
\label{eq:app-phases}
\end{equation}

With these shifts and phases, diagonalizing
\begin{equation}
  (Y_f)_{ij}=\e^{\,p^{f}_{ij}}\Big[1+e^{i\phi_f}\e^{\,\Delta^{f}_{ij}}+e^{i\psi_f}\e^{\,\Delta^{\prime f}_{ij}}\Big]
\end{equation}
and fixing $m_t=169~\mathrm{GeV}$ and $m_b=2.794~\mathrm{GeV}$ at $M_Z$ yields quark masses close to the ``Model'' column of Table~\ref{tab:quarkmasses}.

\section{Two-over-two lattice details}
\label{app:kspace}

The two-over-two relations of Table~\ref{tab:twoover} can be visualized as an integer lattice in $(x,y)$ space as described in Sec.~\ref{sec:results}.
Lines of constant $B$-power satisfy $y-x=9n$ and appear as diagonals separated by $\Delta(y-x)=9$.

For individual quark flavors, the integer coordinates $(x'^{(f)},y'^{(f)})$ defined in Eq.~\eqref{eq:k6k7-quarks} provide a compact map of the charge assignments.
The resulting quark flavor points are shown in Fig.~\ref{fig:k6k7-quarks}.



\end{document}